\documentclass[aip,reprint]{revtex4-1}
\usepackage{graphicx}
\usepackage{dcolumn}
\usepackage{bm}

\usepackage[utf8]{inputenc}
\usepackage[T1]{fontenc}
\usepackage{mathptmx}
\usepackage{etoolbox}
\usepackage{amsmath}
\usepackage{caption}
\usepackage{subcaption}

\draft 

\begin{document}


\title{Shock Wave Interaction with a Naturally Generated Vortex Ring} 



\author{Swapnil Ashok Ahire}
\email[]{swapnilahire15@iitb.ac.in}
\affiliation{Department of Aerospace Engineering, Indian Institute of Technology Bombay, Mumbai, India}
\author{Avijit Chatterjee}
\email[]{avijit@aero.iitb.ac.in}
\affiliation{Department of Aerospace Engineering, Indian Institute of Technology Bombay, Mumbai, India}

\date{\today}

\begin{abstract}
The vortex ring is usually analytically modelled in computational studies involving vortex ring and shock wave interaction. However, incompressibility assumptions of the isolated vortex ring model mostly limits its applicability to the lower Mach number regime. To overcome this limitation, a compressible vortex ring is generated using a pulse jet exiting a circular nozzle, and its interaction with a shock wave is analysed. The shock vortex interaction and resulting sound wave generation is simulated by solving the compressible axisymmetric Navier-Stokes equations. A characteristics-based filter is used to isolate acoustic and hydrodynamic disturbances in the flow-field. A reconfigured computational domain is used to negate the after-effects of the nozzle on the flow field. The strong interaction of compressible vortex rings at higher Mach numbers with different shock wave Mach numbers in the flow field is studied including resultant acoustic wave generation.

\end{abstract}

\pacs{}

\maketitle 

\section{Introduction}

In imperfectly expanded supersonic jets exiting a nozzle, shock waves are present in the jet core and vortex rings are formed near the nozzle lip in the shear layer. The interaction of these vortex rings with shock waves can produce high amplitude noise, such as broadband shock noise and jet screech. This noise in an aircraft can cause damage to the vehicle itself due to structural fatigue and also contribute to community noise. In a supersonic jet, vortex rings are generated at the lip of the nozzle, convect downstream and interact with the shock waves at the edge of the jet core. Multiple interactions of vortex rings and shock waves occur concurrently. The flow field generated due to such interaction can be complex to analyse as a whole. To study this interaction of a vortex ring with a shock wave in a supersonic jet, the problem is often decomposed into a simpler problem, such as the interaction between an isolated vortex ring and a shock wave. 

Sound waves are generated in interactions between columnar vortex and shock wave was reported in experimental studies performed with a shock tube by Hollingsworth\cite{hollingsworth1955schlieren}. In the experiment, a vortex is generated using an airfoil placed inside the shock tube at an angle of attack. In a similar experimental set-up, Dosanjh\cite{dosanjh1965interaction} observed the presence of multiple compression and rarefaction waves with the help of vertical knife edge Schlieren. Pressure distribution of acoustic waves surrounding the vortex can be represented by the superposition of quadrupole, dipole and source terms \cite{dosanjh1965interaction}. Analysis by Ribner\cite{ribner1985cylindrical} also showed that presence of cylindrical acoustic wave.
The linear theory used in the analysis is applicable for weak shock-vortex interactions\cite{ribner1985cylindrical}. Strong interactions which is the focus of the present study involve significant deformation of the shock wave including the formation of secondary shock waves in the shock-vortex interaction\cite{inoue2000successive}.
 
The above-mentioned early experiments were based on columnar vortex and shock interaction, and the first attempt to study vortex ring and shock interaction was reported by Minota\cite{minota1993interaction}. The vortex ring was generated by shock diffraction with a shock tube, and a shock wave propagated by a shock tube in the opposite direction. The Schlieren results showed shock wave deformation and reflection when interacting with a vortex ring. 

Most of the studies performed later on to study vortex shock interaction are computational in nature involving analytically modelled isolated 2D vortex and 3D vortex ring with past experiments and linear analysis serving as an initial validation for the computational set-up. When a 2D vortex is modelled, the total circulation is usually zero \cite{inoue1999sound}, but it is non-zero in axisymmetric 3D vortex ring models \cite{inoue2000successive, meadows1997study}. 
Compressible two-dimensional (2D) Euler equations were solved computationally to study the interaction between a 2D vortex with both weak and strong shock waves by Ellzey and Henneke\cite{ellzey1995interaction}.
In these computations \cite{ellzey1995interaction}, it was observed that a precursor wave initially develops when a shock wave interacts with the flow field surrounding the vortex and expands radially. As the shock wave further interacts with the vortex core, it reflects and moves radially outward, interacting with the precursor. 
Direct numerical simulations\cite{inoue1999sound} were performed on a two-dimensional domain to study shock vortex interaction involving a single vortex and pair of vortices. In this study \cite{inoue1999sound}, it was found that there is negligible effect of Reynolds number, which is defined based on vortex radius on the near-field physics. Near-field physics was found to be dependent on the strength of the vortex and shock wave.

Transition from regular reflection to Mach reflection is observed for a shock wave with higher Mach numbers while interacting with a vortex \cite{ellzey1995interaction,inoue1999sound,inoue2000successive}.
Chatterjee\cite{chatterjee1999shock} proposed a simple model to explain a change in reflection of the shock wave from regular reflection to Mach reflection based on vortex strength and shock strength with inviscid simulations. In Ref.\cite{chatterjee2008multiple}, two-dimensional Euler equations were solved for an extended time using ninth-order weighted essentially non-oscillatory (WENO) schemes for spatial discretization and a fourth-order Runge-Kutta method for time advancement. Previous research had mainly focused on acoustic waves generated immediately after the vortex and shock wave interaction. Results in Chatterjee and Vijayaraj \cite{chatterjee2008multiple} showed that an elliptical rotating vortex also radiates acoustic waves while convecting downstream. 
Shock vortex interaction was also studied on microscale with vortex radius in order of shock wave thickness by Xiao \cite{xiao2014computational}.
The shock wave and vortex interaction is studied in Schardin's problem as well \cite{halder2013numerical, ejtehadi2020investigation}.
Halder et al. numerically studied shock vortex interaction in Schardin's problem and discussed the evolution of acoustic waves along with emergence of diverging acoustic wave. Ejtehadi et al. \cite{ejtehadi2020investigation} developed discontinuous Galerkin method to simulate shock vortex interaction in dusty environment and studied shock vortex interaction along with effect of dust on vortex and shock dynamics.
Kundu and Bisawas \cite{kundu2022analysis}, numerically studied interactions of 2D vortex of higher Mach numbers and a strong shock wave and based on interaction strength, several types the vortex deformation were reported.
Since the majority of the previous studies were conducted with isolated 2D vortex, Skews \cite{skews2021experiments} recently studied interaction of naturally generated vortex and shock waves. For strong interactions, a pressure spike was observed in proximity to the intersection of the reflected and incident shock waves which was not captured in previous studies involving analytically modeled 2D vortex.

There are several computational studies in which 3D vortex ring was modelled analytically for an axisymmetric flow \cite{inoue2000successive,meadows1997study,ding2001computational}. 
The computational study by Meadows \cite{meadows1997study} used the Essentially Non-oscillatory (ENO) method to solve the axisymmetric Euler equations and investigate the interaction between a vortex ring and a shock wave. In this study, 3D vortex ring radius is larger due to the limitations of the analytical model used for vortex ring modelling. It also established the relationship between shock strength and the sound pressure level. 
Inoue and Takahashi\cite{inoue2000successive} carried out a Direct Numerical simulation of SVI (shock-vortex ring interaction) and the propagation of sound waves is studied in more detail.
A different analytical model\cite{inoue2000successive} of the vortex ring used enabled the study of the vortex ring in proximity to the axis. It was observed that repetitive interaction nested in nature with the vortex ring generates multiple compression and rarefaction waves. The variation in the number of sound pressure waves generated in the SVI with vortex Mach number was also presented. Qualitatively the near-field physics of SVI with 2D vortex \cite{inoue1999sound} and 3D vortex ring \cite{inoue2000successive} were found to be similar, though the precursor is found to be weak in the case of the 3D vortex ring.

In numerical simulations which have been employed to examine the interaction between an axisymmetric vortex ring and a shock wave\cite{inoue2000successive,meadows1997study,ding2001computational}, the investigations generally involve analytical modelling of the vortex ring. In-built assumptions, such as incompressibility, are usually present in vortex ring modelling. These models are suitable for SVI studies for vortex rings at relatively lower Mach numbers. 
However, compressibility effects in the vortex ring model become significant in stronger interactions involving higher Mach numbers, such as those encountered in supersonic jets. Though compressible 2D vortex can be modelled analytically \cite{ellzey1995interaction, inoue1999sound,chatterjee1999shock,chatterjee2008multiple} but there is lack of realizable models for an isolated compressible vortex ring to be imposed in numerical simulations for SVI.

\begin{figure}
	\centering
	\begin{subfigure}[b]{0.48\textwidth}
		\centering
		\includegraphics[width=\textwidth]{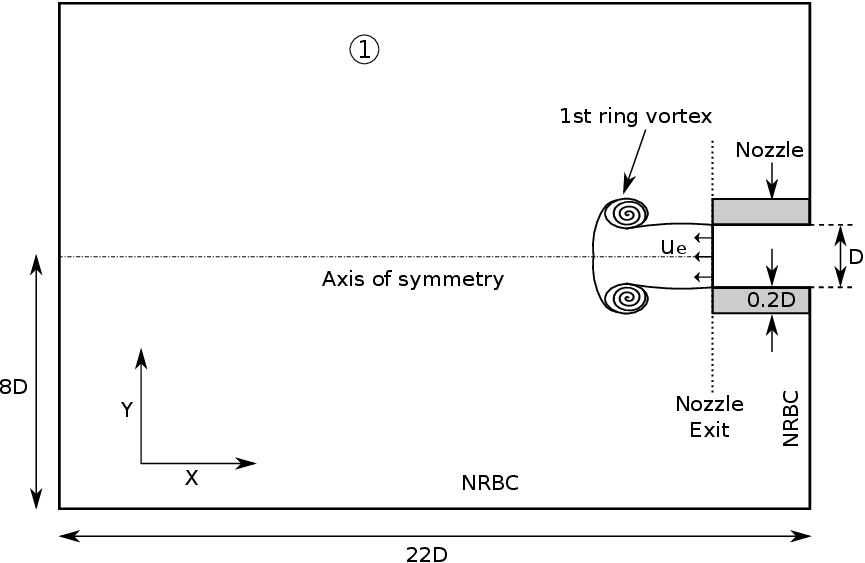}
		\caption{Generation of a vortex ring}
		\label{fig:scheme1}
	\end{subfigure}
	\hfill
	\begin{subfigure}[b]{0.46\textwidth}
		\centering
		\includegraphics[width=\textwidth]{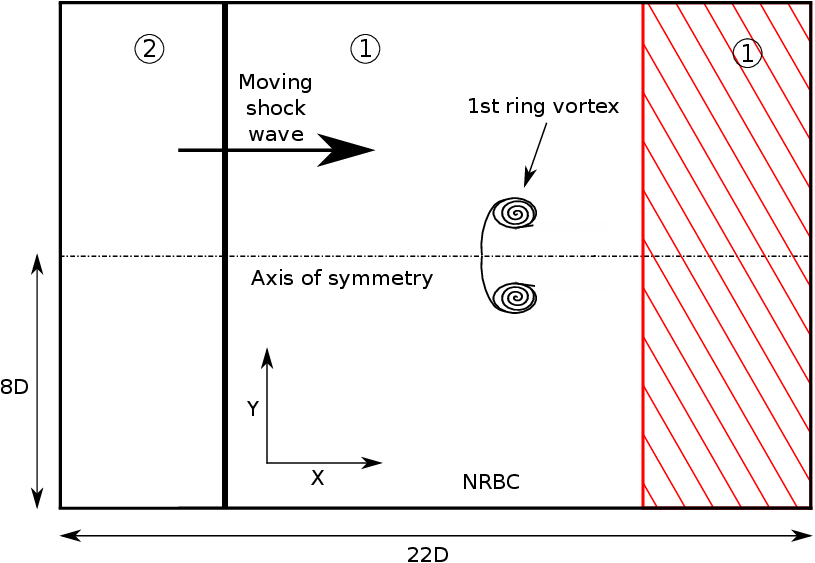}
		\caption{Reconfigured domain}
		\label{fig:scheme2}
	\end{subfigure}
	\caption{Schematic of configuration vortex ring-shock wave for interaction and computational domains}
	\label{fig:schematic}
\end{figure}

The present study uses a high-speed subsonic jet pulse to create a vortex ring. A naturally formed compressible vortex ring can be generated based on the jet Mach number \cite{zhang2019mixing}. The interaction of a compressible vortex ring and moving shock wave is used to investigate the sound waves generated from the interaction. Since the simulations are not performed on isolated vortex rings, there is a possibility of contaminating the set-up with hydrodynamic fluctuations and non-related SVI fluctuations present because of the vortex ring generation process. 
Characteristic-based filters (CBF) \cite{kopitz2005characteristics} are employed to isolate sound pressure and hydrodynamic pressure fluctuations in the flow-field. 
A reconfigured computational domain is used to negate any affect of the nozzle in the flow after the vortex generation process.
Simulations are performed in two parts that is, i) Vortex ring is naturally generated using a jet pulse emanating from a nozzle as shown in Fig. \ref{fig:scheme1}, and ii) Vortex ring interaction with a shock wave with nozzle removed from the domain.
The initial domain configuration for the moving shock wave and compressible vortex ring is shown in Fig. \ref{fig:scheme2} along with the reconfigured domain. A similar computational set-up was used by Takayama\cite{takayama1993self} for studying local jumps in pressure and density at the centre of the vortex ring when it interacted with a moving shock wave.

\section{Numerical Methodology}

The unsteady compressible axisymmetric Navier-Stokes equations are solved to model the SVI. The in-house code uses fifth-order weighted essentially non-oscillatory (WENO) schemes \cite{shu1999high} for spatial discretization, and time advancement is by a Total Variation Diminishing (TVD) second-order Runge-Kutta method similar to Gottlieb \cite{gottlieb1998total}. Higher order WENO discretization schemes are used for studying near-field acoustic flow fields in supersonic jets such as screech \cite{chatterjee2009screech} and are suitable for studying acoustic wave generation process from vortex rings and shock interaction. Non-reflective boundary conditions (NRBC) \cite{jorgenson2002computing} are imposed on domain boundary as shown in the Fig. \ref{fig:schematic}. The nozzle diameter (D) and nozzle lip thickness (0.2D) is also shown in Fig. \ref{fig:schematic}.
The domain is extended 22D and 8D in axial (X) and radial (Y) directions, respectively.

\subsection{Numerical set-up}
\begin{figure}
	\centering
	\includegraphics[width=.45\textwidth]{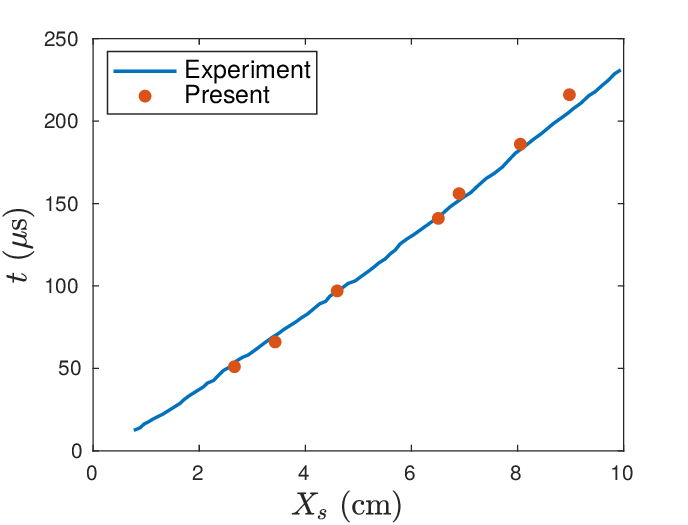}
	\caption{Position of shock wave with time compared with experiments by Ishii \cite{ishii1999experimental}}
	\label{fig:valis}
\end{figure}

The current SVI study involves two distinct parts, a moving shock wave, and a vortex ring generated with a high-speed jet pulse. The validation exercise covers both these parts.
Simulations on shock tube for similar conditions as of that Ishii \cite{ishii1999experimental} at Mach number (M) 1.56 were performed. The position of the shock wave at several time intervals is recorded and validated with the experimental results from Ishii \cite{ishii1999experimental} as shown in Fig. \ref{fig:valis}. 

\begin{figure}
	\centering
	\includegraphics[width=.45\textwidth]{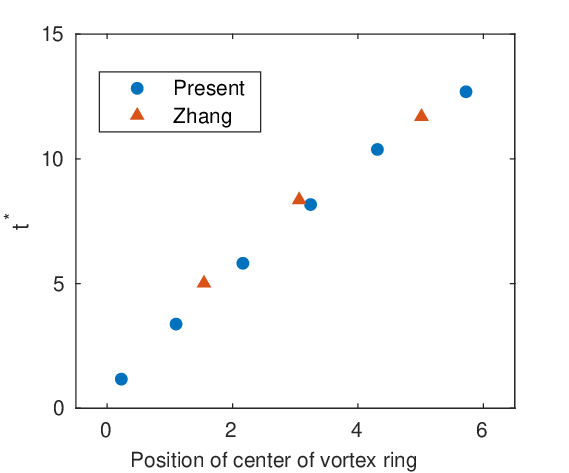}
	\caption{Convection of vortex ring for $M_e = 0.6$ compared with numerical studies by Zhang \cite{zhang2019mixing}}
	\label{fig:valiv}
\end{figure}

In simulations conducted by Zhang \cite{zhang2019mixing}, a vortex ring evolves at the nozzle lip due to a high-speed subsonic jet emanating from a nozzle at Mach number ($M_e$) = 0.6 and Reynolds Number, $Re_D = 1.3 \times 10^5$. The code was validated for the evolution of the vortex ring at $M_e = 0.6$, and one of the results from the validation is shown in Fig. \ref{fig:valiv}, which illustrates the convection of the vortex ring over time. In some studies involving transonic jet ($M_e$ = 0.9) such as by Arnaud et al. \cite{fosso2015subsonic}, temperature profile is imposed at the exit of a nozzle with velocity profile using Crocco-Buseman relation. In the study, effect of density variation is studied for $M_e$ = 0.6. No significant changes were observed when the vortex ring center was traced over time. 

\begin{figure}
	\includegraphics[width=.45\textwidth]{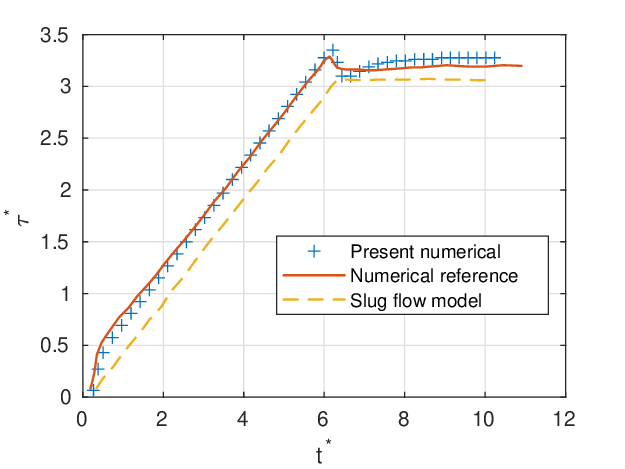}
	\caption{Validation for unsteady pulse of velocity at nozzle exit for $M_e = 0.2$ for $L/D = 6$ with Zhao et al. \cite{zhao2000effects}}
	\label{fig:unseady_valid}
\end{figure}

When a vortex ring is generated naturally, the primary vortex ring is accompanied by a trailing jet. Over time, this trailing jet becomes unstable, leading to the formation of secondary vortex rings \cite{zhao2000effects}. To circumvent the possibility of secondary vortex ring generation, a velocity pulse is created at the nozzle exit instead of continuous flow. This pulse is calibrated to be large enough to generate the primary vortex ring, but small enough to prevent the formation of secondary vortex rings. The velocity pulse imposed is explained in detail in Section \ref{initial}.   
The code is also validated for the primary vortex ring generated with an unsteady pulse of velocity at the nozzle exit. 
Zhao et al. \cite{zhao2000effects} numerically studied the effects of an impulsively started jet and its impact on trailing jet instabilities. Numerical simulations were performed at $M_e = 0.2$, modeling various L/D ratios, where L is the distance traveled by the piston to force the fluid through a nozzle of diameter D. These simulations corresponded to the experiments conducted by Gharib et al. \cite{gharib1998universal}.
Under similar conditions for $L/D = 6$, numerical simulations are performed. The evolution of the vortex ring, in terms of total circulation over time, is shown in Fig. \ref{fig:unseady_valid}. Both circulation and time are non-dimensionalized using the maximum velocity imposed at the nozzle and the nozzle diameter. Total circulation ($\tau^*$) initially increases with time and then remains constant as time progresses, as illustrated in Fig. \ref{fig:unseady_valid}.

\subsection{Initial Conditions} \label{initial}
\begin{figure}
	\centering
		\includegraphics[width=.4\textwidth]{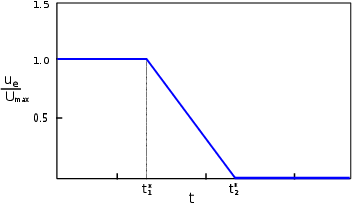}
        \caption{Velocity pulse imposed at nozzle exit}
	\label{fig:unsteady_v}
\end{figure}

The simulations are performed in two parts. The first part involves the generation of a primary vortex ring using a velocity pulse imposed at the nozzle. The results from the first part serve as the initial conditions for the second part, which is to study the interaction between a vortex ring and a moving shock wave without the presence of the nozzle.
At the nozzle exit, subsonic flow conditions are applied with fully expanded pressure ($P = P_1$), where $P_1$ is the ambient pressure similar to Zhang et al \cite{zhang2019mixing}. The exit axial velocity of the jet from the nozzle $u_e$, is based on the jet Mach number ($M_e$). The inflow velocity profile is steady and laminar, similar to that used by Zhang et al. \cite{zhang2019mixing}. In the present study, vortex rings are generated for three different Mach numbers as shown in Table \ref{tab:table2} and Reynolds number $Re_{D}$ corresponding to maximum pulse velocity. At the nozzle exit, a velocity pulse is imposed as depicted in Fig. \ref{fig:unsteady_v}. The velocity profile for the pulse is given by:

\begin{equation}{\label{eq:unsteady}}
	\frac{u_e}{U_{max}}=\begin{cases}
		1, &  0 \le t \le t_1^*\\
		mt + y_c, &  t_1^* < t \le t_2^*  \\
		0, & t > t_2^*
	\end{cases}
\end{equation}
where,
\begin{equation*}
	 m = -1/(t_2^* - t_1^*) \text{, and }
	 y_c = -mt_2^*.
\end{equation*}

\begin{figure}
	\includegraphics[width=.48\textwidth]{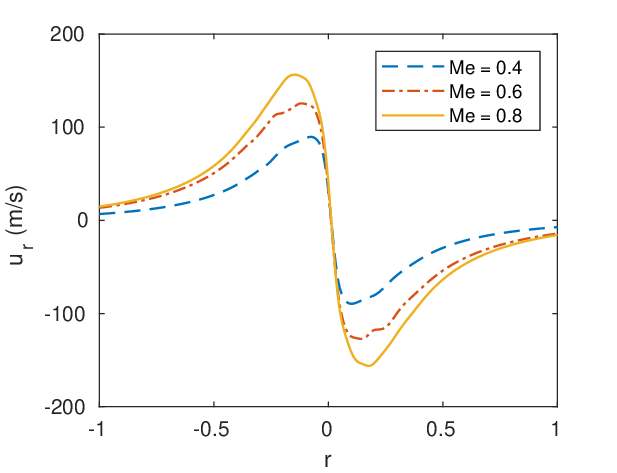}
	\caption{Radial velocity $u_r$ distribution in axial directions from the center of vortex ring cross-section}
	\label{fig:velocity_y}
\end{figure}

The values of $t_1^*$ and $t_2^*$ used in this study are listed in Table \ref{tab:table2}. 
Here, $t_1$ is chosen so as to only generate a primary vortex ring in all the cases while $t_1 - t_2$ is kept constant for all the cases.
Additionally, $U_{max}$ represents the maximum velocity imposed, corresponding to the jet exit Mach number ($M_e$). The radial velocity profile in an axial direction from center of the vortex ring cross-section is shown in Fig. \ref{fig:velocity_y}. 
The vortex ring Mach number ($M_v $) is measured as the maximum tangential velocity $u_{\theta_{max}}$ excluding the convective velocity of the vortex ring. Similar definition of $M_v$ is used in studies involving 2D vortices such as those by Inoue et al. \cite{inoue1999sound} and Chatterjee et al. \cite{chatterjee2008multiple}. 
The vortex ring Mach number depends on the jet exit Mach number.
In-principle stronger vortex rings can be generated to study SVI till the jet exit Mach number turns supersonic and exit flow is no longer shock free resulting in shocklets embedded in the generated vortex \cite{ishii1999experimental}.

In Fig. \ref{fig:unseady_valid}, velocity pulse results in an increase in circulation after which it is invariant. Due to the imposed momentum pulse, the vortex ring continues to convect away from the nozzle. Once the vortex ring has moved sufficiently far from the nozzle exit, the domain is truncated near the nozzle to negate any after-affects of the nozzle on the flow-field including later interaction with the moving shock-wave.
The truncated part of the domain is re-initialized with conditions corresponding to $P_1$.
This modified domain conditions are then used as the initial condition for the second part of the simulation. In this phase, a moving shock wave of strength ($M_s$), traveling from left to right, is generated, as illustrated in Fig. \ref{fig:scheme2}.

\begin{table}
	\caption{\label{tab:table2} Parameters used for generation of vortex ring, $t^* = t c_{1} / D $, $M_v = u_{\theta max} / c_{1}$ }
	\begin{ruledtabular}
		\begin{tabular}{ccccc}
			$M_e$ & $t_1^*$&$t_2^*$ &$M_v$ & $Re_{D} \times 10^{5}$\\
			\hline
			0.4& 3.4 & 5.1  & 0.26 & 0.96 
			 \\
			0.6& 3.0 & 4.7 & 0.37 & 1.45 
			\\
			0.8 & 1.7 & 3.4 & 0.46 & 1.93

		\end{tabular}
	\end{ruledtabular}
\end{table}

\subsection{Grid independence}
\begin{figure}
	\centering
		\includegraphics[width=.45\textwidth]{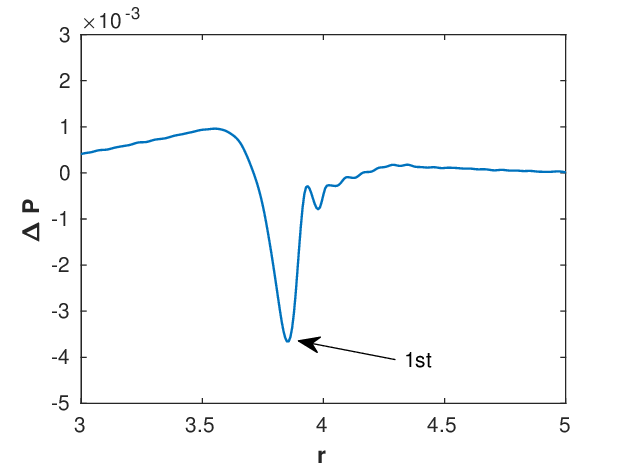}
	\caption{Acoustic waves generated along the jet axis post SVI at $\bar{t} = 4.32$, where $\bar{t} = t c_{2}/D$}
	\label{fig:gis_wave}
\end{figure}

\begin{figure}
		\includegraphics[width=.45\textwidth]{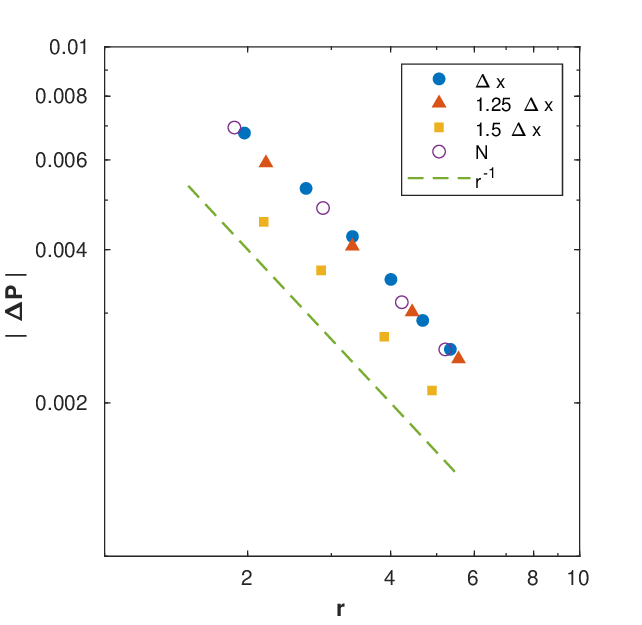}
		\caption{Decay of magnitude of $\Delta P$ for 1st trough and r is the distance from center of the vortex ring to trough}
        \label{fig:gis1}
\end{figure}

Grid independence studies are performed to make sure that the solution is not dependent on the underlying discretization.
Uniformly distributed grid points are used for the present study. In order to achieve grid-independent solutions, SVI simulations for case A, as shown in Table \ref{tab:svi}, were performed for three grid configurations: I. $\Delta x_1 = 1.07 \times 10^{-2}$, II. $\Delta x_2 = 1.25 \times \Delta x_1$ and III. $\Delta x_3 = 1.5 \times \Delta x_1$, where all lengths are non-dimensionalized with the diameter of the nozzle (D), $\Delta x$ is the grid size in an axial direction and aspect ratios of the grid cells are constant. The number of grid points in the x and y direction for grid I are $3071 \times 1000$, for grid II are $2300 \times 750$, and for grid III are $1532 \times 500$. In the numerical simulation, pressure fluctuations are generated along the axis of the subsonic jet due to SVI, as seen in Fig. \ref{fig:gis_wave}, where the leading first pressure fluctuation is labelled as such (1st) for case A.
For the grid independence study, case A, was chosen because the amplitude of the sound pressure is relatively lower compared to other cases considered, as shown in the study by Meadows \cite{meadows1997study}, where the dependence of sound pressure on the shock wave Mach number is demonstrated. A grid capable of resolving the least magnitude of sound pressure of case A would ensure that the sound pressure waves of higher magnitude are resolved. 
As in previous studies, sound pressure ($\Delta P$) is computed by non-dimensionalizing pressure with the pressure behind the shock wave ($P_2$) and defined as $\Delta P = (P-P_2)/P_2$. For three grid configurations, 1st pressure fluctuation as shown in Fig. \ref{fig:gis_wave} at several positions are plotted in Fig. \ref{fig:gis1}. 
By examining Fig. \ref{fig:gis1}, the sound pressure for the relatively coarsest grid shows a small deviation, but grid I and grid II are a good match. Along with grid independence with sound pressure, the number of sound waves generated are also found to be invariant with the grid levels.
For all grids, a constant time step corresponding to Courant Freidrich Lewy (CFL) number is used. Based on grid independence study, grid I is used for all the cases shown in Table \ref{tab:svi}.

In order to evaluate the effect on SVI due to re-initialization of the domain conditions, simulations were also performed for case A with a nozzle present (without re-initialization) on grid I and shown as $N$ in Fig. \ref{fig:gis1}. There is an insignificant effect of domain re-initialization on the magnitude of pressure waves generated, as evinced by Fig. \ref{fig:gis1}. 
This can be explained by vorticity distribution near the vortex ring. 
The vorticity and circulation is stronger in the area near the vortex ring and tends to become negligible relatively far away from the vortex ring center. 
This property of the vorticity distribution is exploited while re-initializing the domain. In simulations with the nozzle, it was observed that the shock wave is reflected off the nozzle in a manner similar to the interaction between a square cavity and a planar shock, as reported by Igra et al. \cite{igra1996experimental}. Therefore, there is a possibility of contamination from the waves reflected if the nozzle is not sufficiently far away from the vortex ring. In this study, nozzle is removed due to re-initialization of the domain, therefore, apart from SVI, it minimizes the presence of extraneous noise generation.

\section{Results and Discussion}

\begin{figure*}
	\centering
	\begin{subfigure}[b]{0.30\textwidth}
		\centering
		\includegraphics[width=\textwidth]{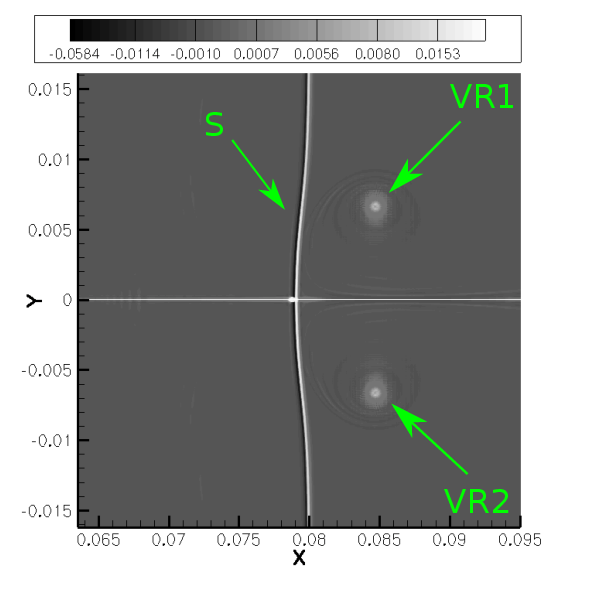}
		\caption{$\bar{t} = -0.16$}
		\label{fig:40}
	\end{subfigure}
	\hfill
	\begin{subfigure}[b]{0.3\textwidth}
		\centering
		\includegraphics[width=\textwidth]{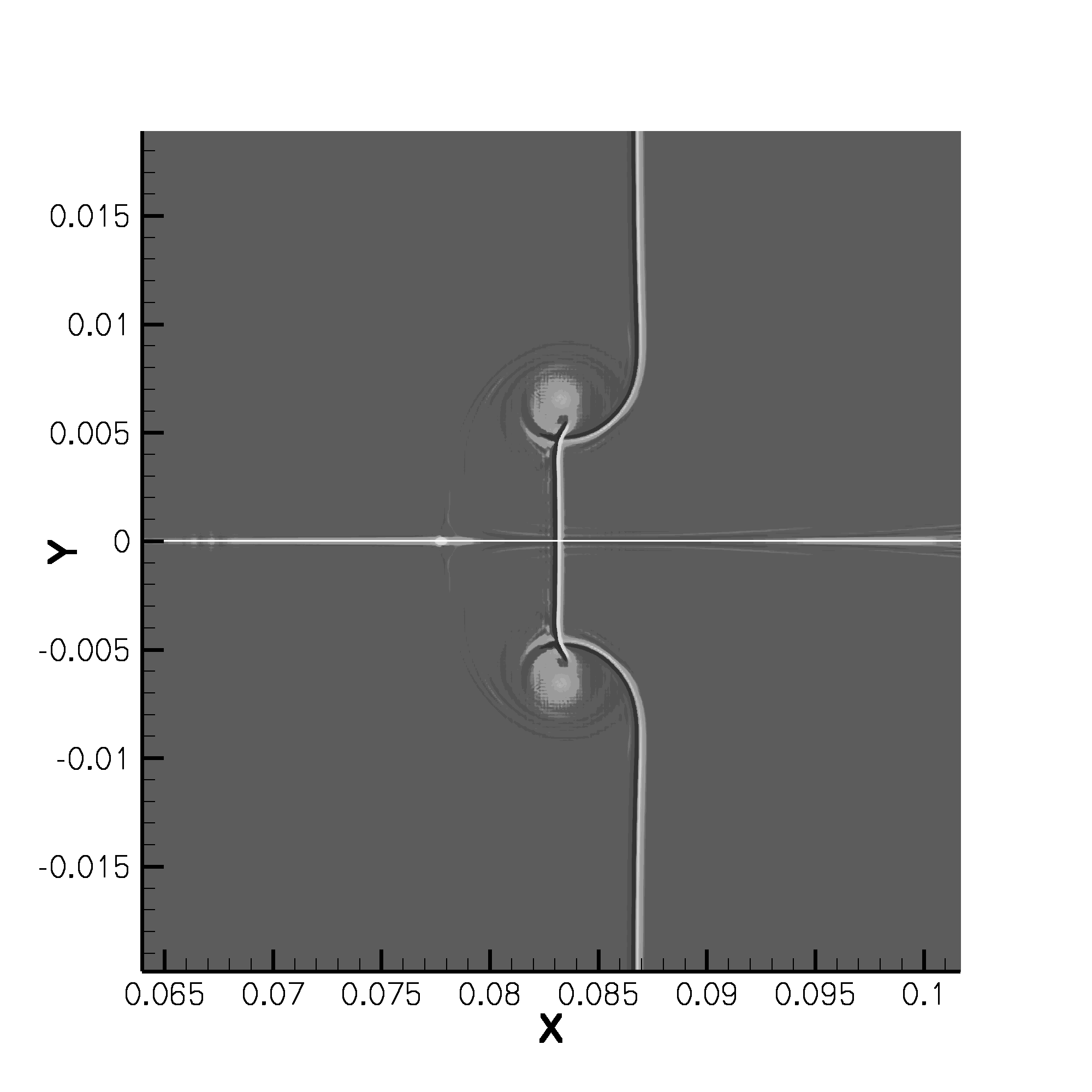}
		\caption{$\bar{t} = 0.49$}
		\label{fig:44}
	\end{subfigure}
	\hfill
	\begin{subfigure}[b]{0.30\textwidth}
		\centering
		\includegraphics[width=\textwidth]{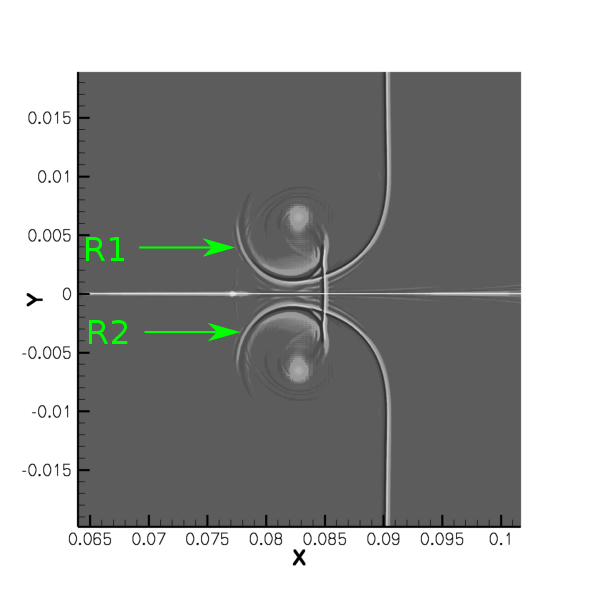}
		\caption{$\bar{t} = 0.82$}
		\label{fig:46}
	\end{subfigure}

	\begin{subfigure}[b]{0.3\textwidth}
		\centering
		\includegraphics[width=\textwidth]{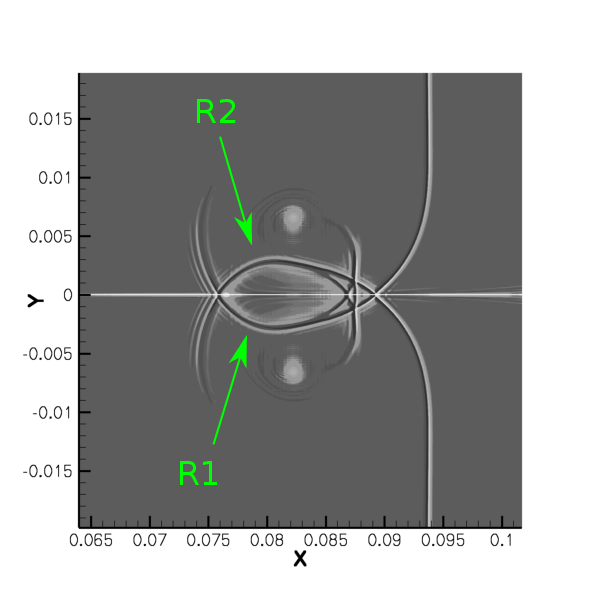}
		\caption{$\bar{t} = 1.16$}
		\label{fig:48}
	\end{subfigure}
	\hfill
	\begin{subfigure}[b]{0.3\textwidth}
	\centering
	\includegraphics[width=\textwidth]{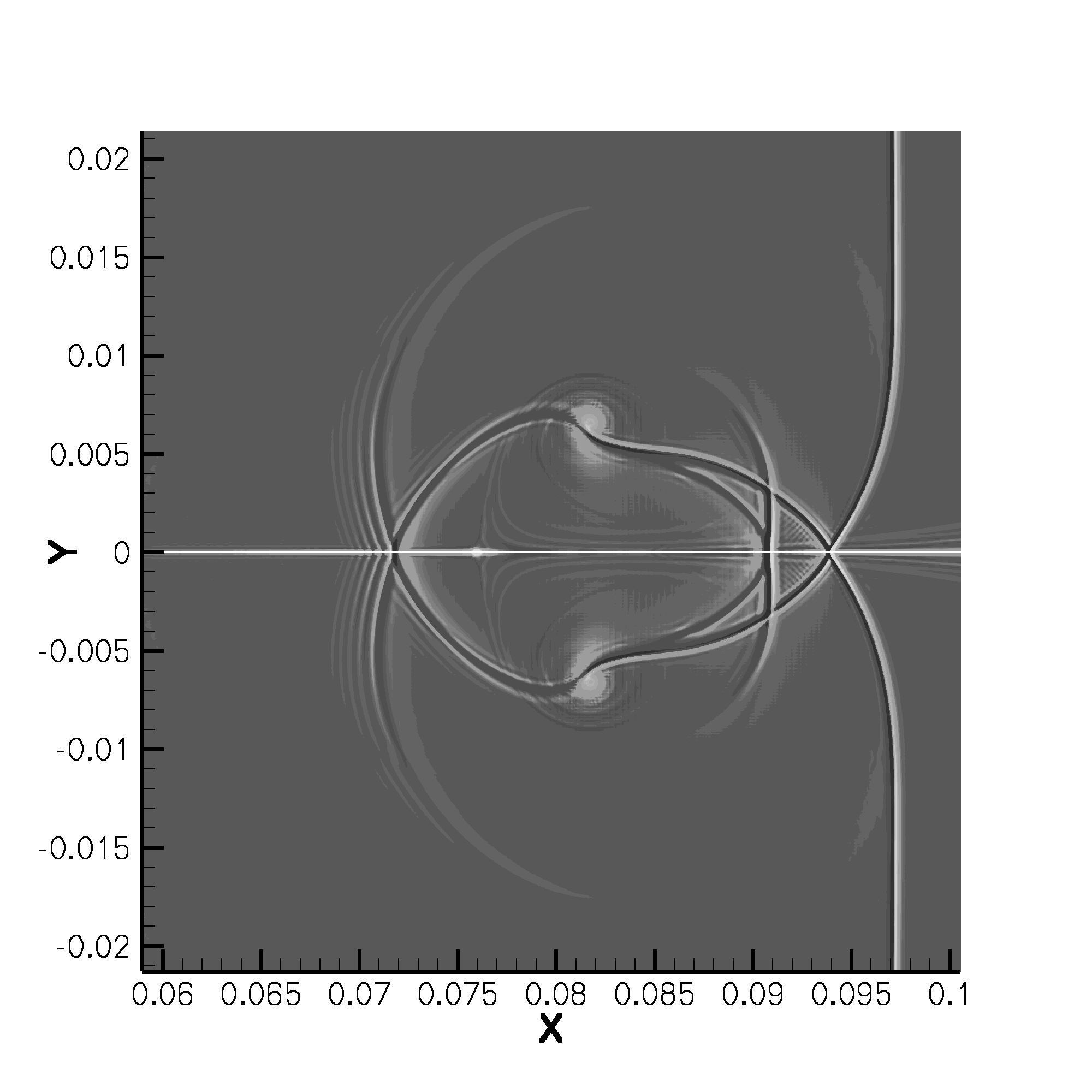}
	\caption{$\bar{t} = 1.50$}
	\label{fig:50}
	\end{subfigure}
	\hfill
	\begin{subfigure}[b]{0.3\textwidth}
		\centering
		\includegraphics[width=\textwidth]{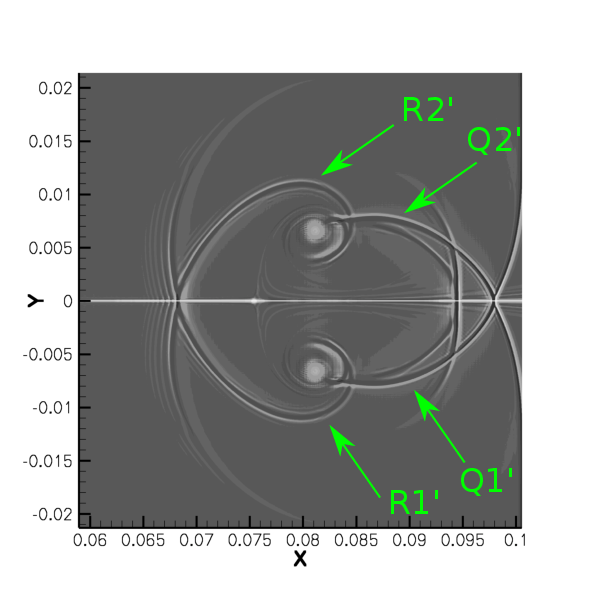}
		\caption{$\bar{t} = 1.83$}
		\label{fig:52}
	\end{subfigure}
	\begin{subfigure}[b]{0.3\textwidth}
	\centering
	\includegraphics[width=\textwidth]{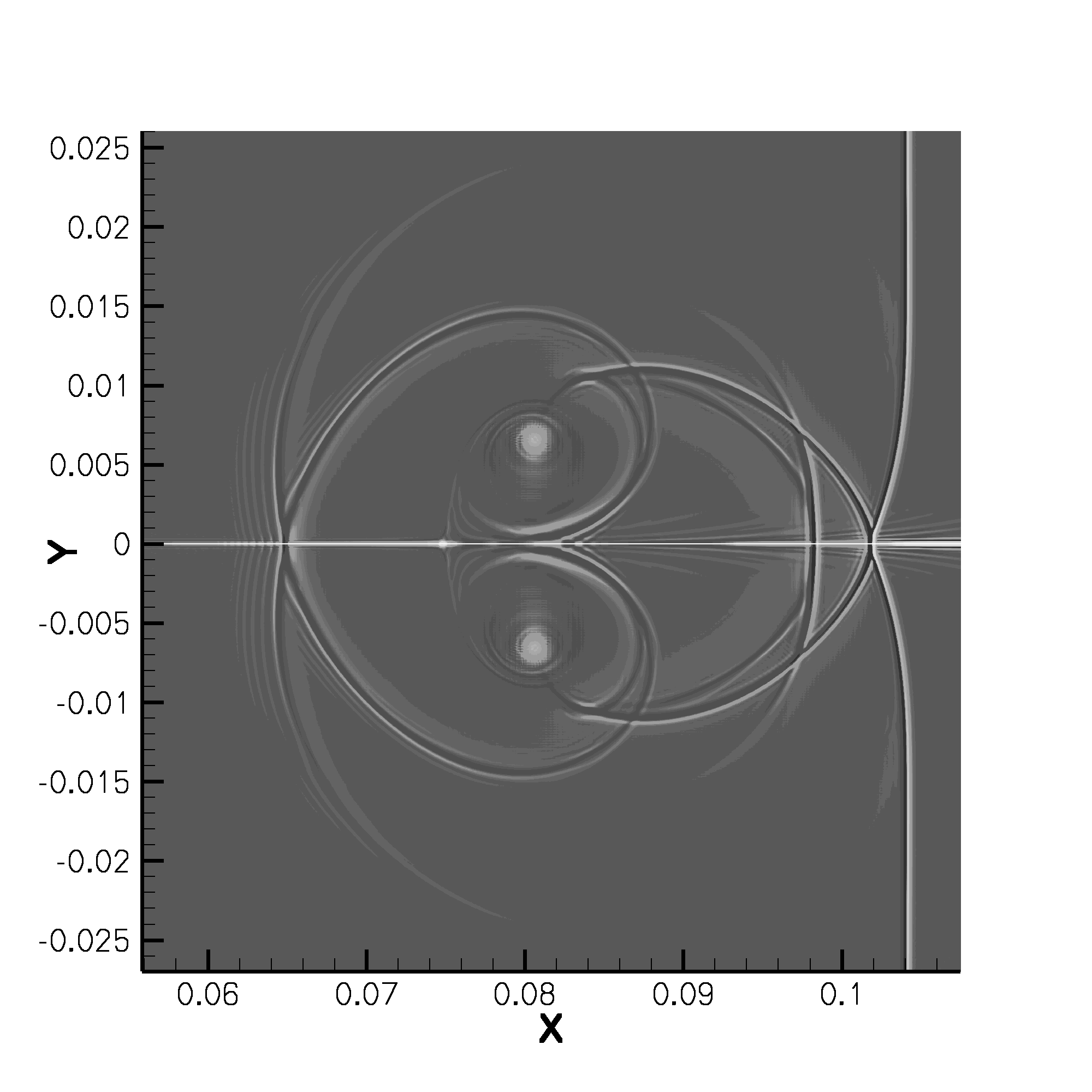}
	\caption{$\bar{t} = 2.17$}
	\label{fig:54}
	\end{subfigure}
	\hfill
	\begin{subfigure}[b]{0.3\textwidth}
	\centering
	\includegraphics[width=\textwidth]{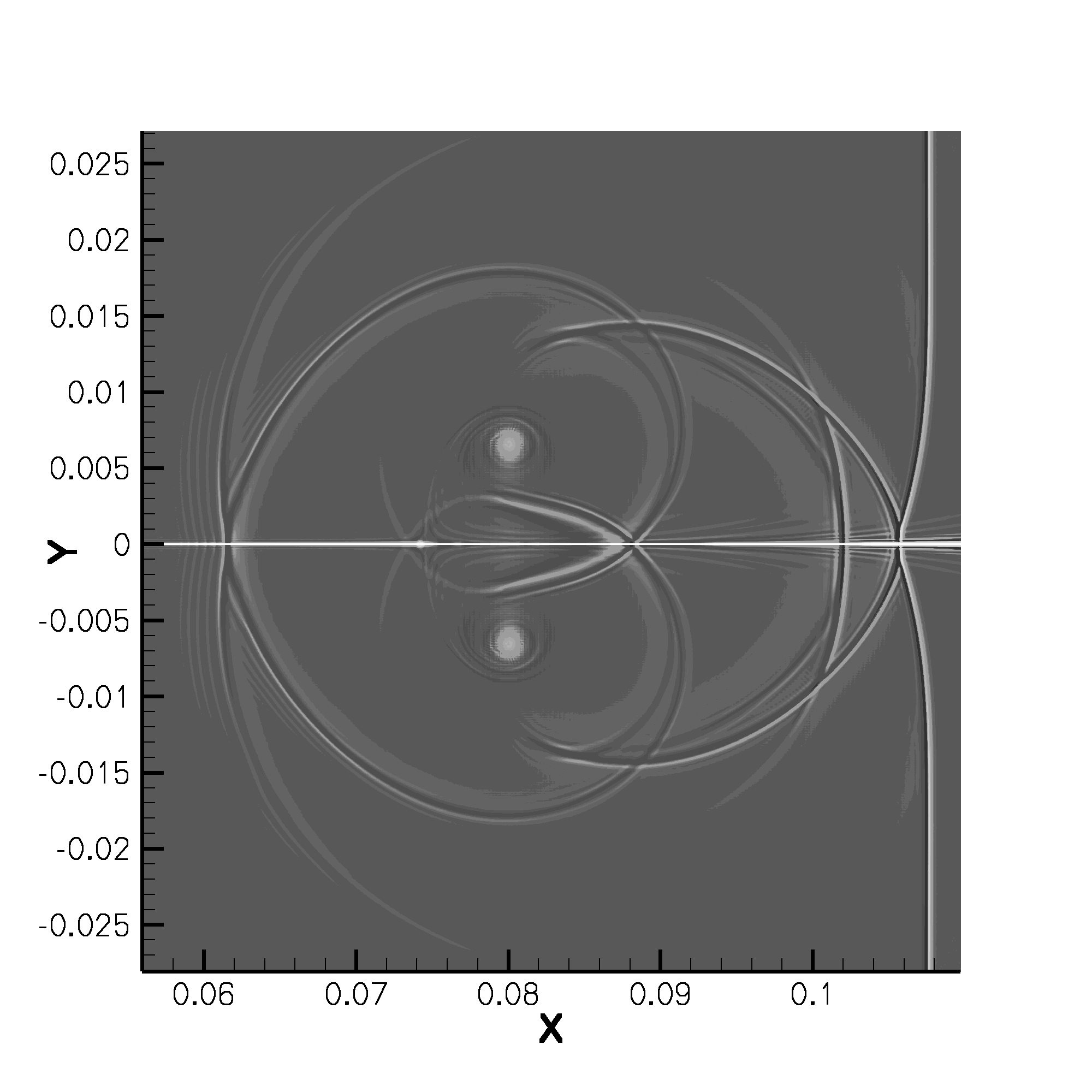}
	\caption{$\bar{t} = 2.50$}
	\label{fig:56}
	\end{subfigure}
	\hfill
	\begin{subfigure}[b]{0.3\textwidth}
	\centering
	\includegraphics[width=\textwidth]{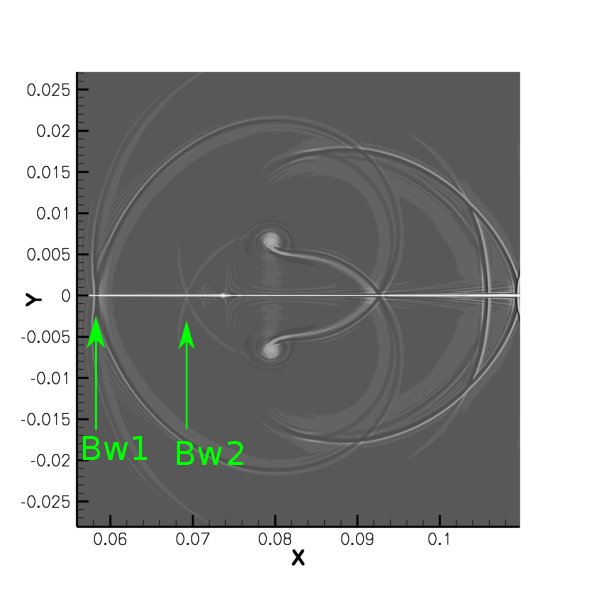}
	\caption{$\bar{t} = 2.85$}
	\label{fig:58}
\end{subfigure}	
	\caption{Numerical shadowgraphs of SVI showing evolution of acoustic pressure waves for case B and $\bar{t} = t c_{2 \infty}/D$}
	\label{fig:shadow}
\end{figure*}

The numerical shadowgraphs at different time instants are presented for the interaction between a clockwise rotating vortex ring generated with a high-speed jet pulse exiting nozzle at Mach number, $M_e$=0.6 and a shock wave with Mach number, $M_s$=1.05, that is case B in Table \ref{tab:svi}. The time evolution of the flow field resulting from the interaction is shown in Fig. \ref{fig:shadow}. The shock wave propagates from left to right and the naturally generated vortex ring travels from right to left. Vortex ring cross-sections on top and bottom are referred to as VR1 and VR2, respectively, shown in Fig. \ref{fig:40} with shock wave denoted as S.
As the moving shock wave approaches the vortex ring, it deforms, as shown in Figs. \ref{fig:40} and \ref{fig:44}. Once the shock wave passes the vortex ring, two reflected waves are generated around the vortex ring, as shown in Fig. \ref{fig:46}. One of the reflected waves (R1) encircles VR1 and moves towards the jet's axis away from VR1, while the other reflected shock wave is almost parallel to the incident shock wave. The reflected waves (R1 and R2) propagates cylindrically towards the jet's axis as seen in Figs. \ref{fig:44} and \ref{fig:46}. Later, the reflected wave (R1) passes through the jet axis and moves towards VR2, as shown in Fig. \ref{fig:48}. The reflected wave (R1) approaches VR2, as shown in Fig. \ref{fig:48}, and the interaction with VR2 produces rarefaction and compression waves (R1' and Q1') which is evident from Figs. \ref{fig:50} and \ref{fig:52}. This wave (R1') encircles the VR2, while the other wave (Q1') moves away from VR2, as shown in Fig. \ref{fig:54}. R1' again moves towards VR1 crossing the axis of the jet and a new compression wave interacts with VR1 which leads to generation of two waves as viewed in Figs. \ref{fig:54}, \ref{fig:56} and \ref{fig:58}. 
A very similar behaviour of waves generated by the shock wave and isolated vortex ring interaction was observed by Inoue \cite{inoue2000successive} using an analytically modeled vortex ring.
These interactions repeats itself based on strength of the vortex ring and incident shock wave.
These identical and nested interactions are characteristic of a strong SVI involving a ring vortex with the interactions become progressively weaker.

Previous studies have been performed on isolated vortex ring - shock wave interaction and in-principle there are no other physical mechanisms other than SVI to generate fluctuations in the flow.
However, when a vortex ring is generated naturally using a jet emanating from a nozzle, there are chances of hydrodynamic fluctuations in the flow unrelated to SVI but related to vortex generation process.
A characteristics based filter is utilized to isolate fluctuations due to SVI from hydrodynamic fluctuations. It may noted that numerical simulations due to an analytically modeled vortex ring can also contain spurious waves due to truncation and other issues related to numerical implementation of the analytical model in a finite domain \cite{inoue2000successive}.


\begin{table}
	\caption{\label{tab:svi} Parameter used for simulations to study SVI}
	\begin{ruledtabular}
		\begin{tabular}{ccccc}
			Case & $M_v$ & $M_s$ & W \\
			\hline
			A & 0.26 & 1.05  & 1
			\\
			B & 0.37 & 1.05 & 2
			\\
			C & 0.46 & 1.05 & 3
			\\
			D & 0.26 & 1.20 & 1
			\\
			E & 0.37 & 1.20 & 2
		\end{tabular}
	\end{ruledtabular}
\end{table}

\subsection{Characteristics Based Filter}
CBF was proposed by Kopitz \cite{kopitz2005characteristics} primarily to capture acoustic signals in turbulence-dominated flow simulations. 
CBF operates by segregating acoustic and hydrodynamic fluctuations, using the difference in convective speed of these fluctuations. If the convective speed closely aligns with the speed of sound, it signifies that the fluctuations are predominantly of an acoustic nature. This method in the present study enables the isolation and characterization of acoustic phenomena amid possible hydrodynamic fluctuations due to subsonic jet in the computational domain. In the current study, fluctuations due to SVI propagate into reservoir conditions behind the propagating shock wave, making the CBF a useful tool for segregation of acoustic and hydrodynamic fluctuations.

In characteristic theory for one-dimensional and first-order hyperbolic partial differential equations, the two acoustic related characteristics variables along characteristics in linear acoustics are expressed as \cite{leveque2002finite, kopitz2005characteristics}:
\begin{equation}
	f_a = \frac{1}{2} \left(\frac{p_a}{\rho a} + u_a \right)
\end{equation}

\begin{equation}
	g_a = \frac{1}{2} \left(\frac{p_a}{\rho a} - u_a \right)
\end{equation}
Here, $p_a$ and $u_a$ denote fluctuations in pressure and velocity, respectively, with respect to the mean pressure ($p_m$) and mean velocity ($u_m$). Additionally, $\rho$ represents mean density, and $a$ corresponds to the speed of sound. 
The above two characteristics variables define acoustic wave propagation along characteristics, with $f_a$ propagating downstream and $g_a$ travelling upstream in the one-dimensional domain. Characteristics variables retain their shape while propagating along these characteristic lines. Thus, 

\begin{equation}\label{eq:ga}
	g_a(x+\Delta x, t+ \Delta t) = g_a(x,t)
\end{equation}
\begin{equation}\label{eq:dx}
	\Delta x = (a - u_{m}) \Delta t
\end{equation}
where, $\Delta t$ is the time lapse.

\begin{figure}[h]
	\centering
	\includegraphics[width=.45\textwidth]{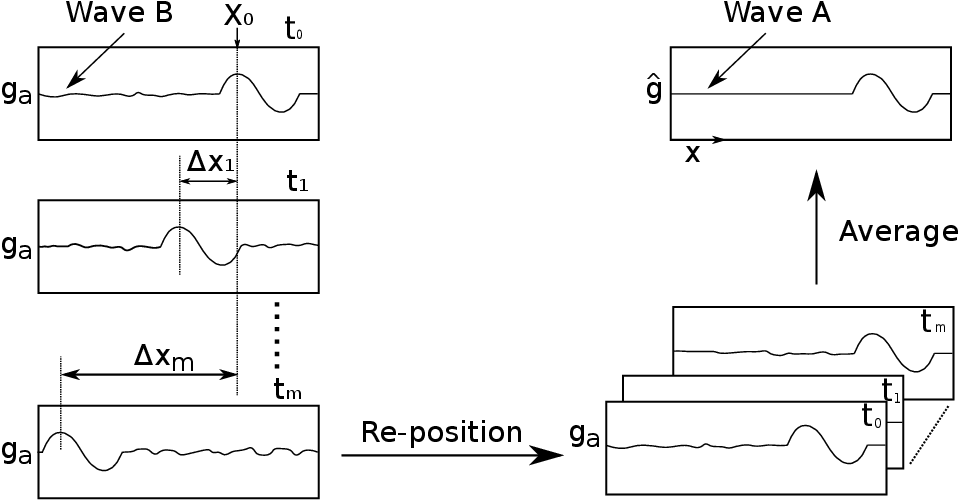}
	\caption{Schematic of Characteristics Based Filter, Wave A is filtered acoustic wave (without non-acoustics fluctuations) $\hat{g}$ and Wave B is characteristics variables $g_a$ at time $t_i$}
	\label{fig:schematic_cbf}
\end{figure}

In Fig. \ref{fig:schematic_cbf}, Wave A is an acoustic wave generated by characteristic based filtering performed on a set of Wave B's at times $t_0, t_1, \dots t_m$, which may contain hydrodynamic fluctuations. The fluctuations in Wave B's are acoustic in nature if they move $\Delta x$ distance computed by equation(\ref{eq:dx}) from any reference position $X_0$. The moving fluctuations in Wave B will be re-positioned with respect to reference position $X_0$ based on computed $\Delta x_i$ as shown in Fig. \ref{fig:schematic_cbf}. 
Wave A will be generated by the mean fluctuations of all the re-positioned waves in Wave B, effectively eliminating most of the hydrodynamic fluctuations.

The following formula can summarize the entire process,
\begin{equation}
	\hat{g} (x,t) = \frac{1}{m+1}  \sum_{i=0}^m g (x-i(a-u_m) \Delta t, t-i \Delta t) 
\end{equation}

\begin{figure}
	\centering
	\begin{subfigure}[b]{0.45\textwidth}
		\centering
		\includegraphics[width=\textwidth]{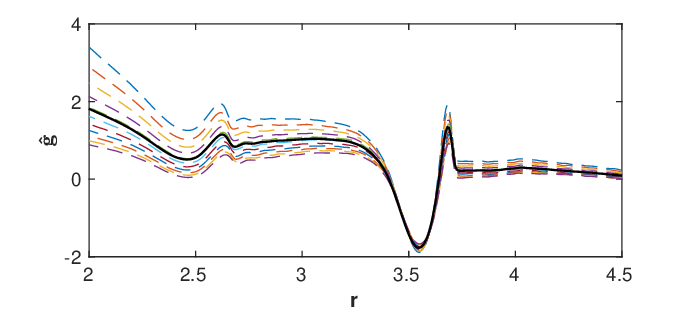}
		\caption{CBF from time $\bar{t_1} = 4.87 \text{ to } \bar{t_n} = 6.55$, ------ : CB filtered result, - - - - : snapshot data after re-positioning the waves as shown in Fig. \ref{fig:schematic_cbf}}
		\label{fig:cbf_g}
	\end{subfigure}
	\hfill
	\begin{subfigure}[b]{0.45\textwidth}
		\centering
		\includegraphics[width=\textwidth]{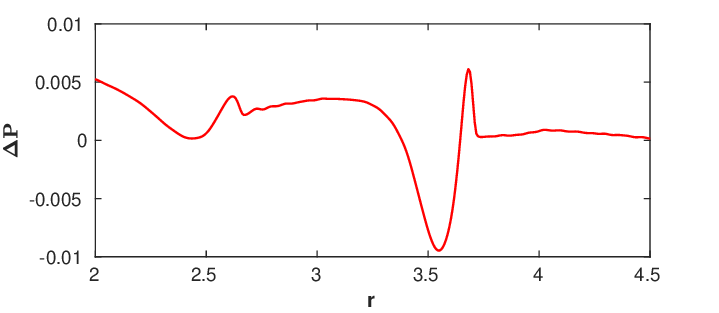}
		\caption{Pressure variation along the axis of the jet at $\bar{t} = 4.87$}
		\label{fig:cbf_p}
	\end{subfigure}
	\caption{CB filtered and instantaneous $\Delta P$ comparison of case B}
	\label{fig:cbf_example}
\end{figure}

In the present study, CBF \cite{kopitz2005characteristics} is used to get acoustic fluctuations along the axis of the jet where this characteristic theory is applicable due to wave fronts moving normal to the axis in axisymmetric flow and can be treated as planar waves. As pressure wave moves away from the vortex ring center, amplitude of the acoustic pressure drops because of spherical spread of the wave. Therefore, small time intervals are chosen for applying CBF, so that the pressure drop can be neglected in the analysis. In the present study, flow quantities required to compute the characteristics variables along the axis of the jet are stored for multiple time instances and are called snapshots. To replicate an acoustic wave, flow quantities for a time-windows are used; for example, in Fig. \ref{fig:cbf_g}, $\bar{t_1}$ is the starting time, and $\bar{t_n}$ is the end time for a given window. In a time window, a set of snapshots can be used for filtering out hydrodynamic fluctuations. In the present analysis, it was found that at least ten snapshots in a window are required to replicate acoustic waves, which do not change with the number of snapshots used, and is sufficient to capture moving acoustic waves. Instantaneous sound pressure ($\Delta P$) along with the axis of the jet is shown in the Fig. \ref{fig:cbf_example} along with filtered characteristics variables $\hat{g}$.  
It is further verified that this pressure fluctuation (and the characteristics variable) travel with the local computed speed of sound.

The vortex ring was generated by a subsonic jet nozzle, raising the possibility of contamination by hydrodynamic fluctuations. However, comparison of filtered and instantaneous results as shown in Fig. \ref{fig:cbf_example}, shows almost no contamination present at these moderate Mach numbers. The figure illustrates the clean formation and propagation of acoustic waves without significant interference from hydrodynamically induced pressure fluctuations. 
The pressure fluctuations due to SVI for lower $M_v$ values, such as in case A, are also consistent with studies involving isolated vortex rings \cite{inoue2000successive}, indicating lack of spurious acoustic variations from the vortex generation process.
This overall lack of contamination ensures the reliability of the simulation results and confirms the effectiveness of the methods used to generation of vortex ring to study SVI. 
However, CBF is always used for processing results as higher Mach numbers of the vortex as a safeguard against possible hydrodynamic contamination of the flow field due to the vortex generating process. 
The filtered characteristic variable $\hat{g}$ obtained from CBF, as also shows above, is interchangeable with the filtered acoustic pressure other than the magnitude as it scales similarly.

\subsection{Effect of $M_v$ and $M_s$}{\label{mv}}

In cases A, B, and C in Table \ref{tab:svi}, the shock wave Mach number ($M_s$) is held constant, while the vortex ring Mach number ($M_v$) is varied to study the impact of $M_v$ on the pressure waves generated due to SVI. 
Each nested interaction involves a reflected shock interacting with the primary vortex, producing an acoustic pressure pulse consisting of a pair of consecutive rarefaction and compression waves, as is also observed in the work of Inoue et al.\cite{inoue2000successive}.
In the current work these pulses are labelled $w_1, w_2, \ldots, w_n$ and it should be possible to connect each pulse with an unique nested interaction in the SVI process.

\begin{figure}[h]
	\centering
	\includegraphics[width=.4\textwidth]{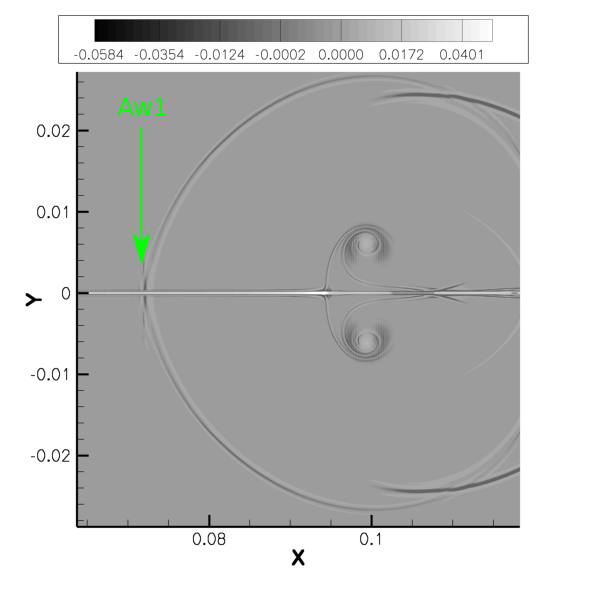}
	\caption{Numerical shadowgraphs of SVI for Case A at $\bar{t} = 3.13$}
	\label{fig:c5_contour}
\end{figure}

\begin{figure}[h]
	\centering
	\includegraphics[width=.4\textwidth]{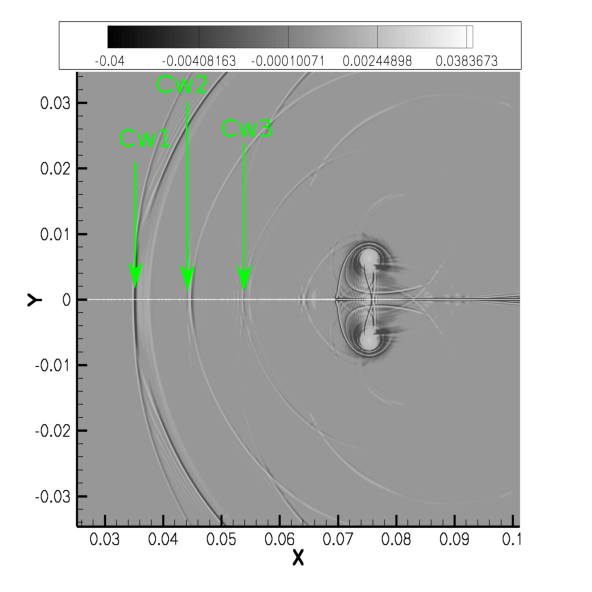}
	\caption{Numerical shadowgraphs of SVI for Case C at $\bar{t} = 5.60$}
	\label{fig:c3_contour}
\end{figure}

\begin{figure}[h]
	\centering
	\includegraphics[width=.45\textwidth]{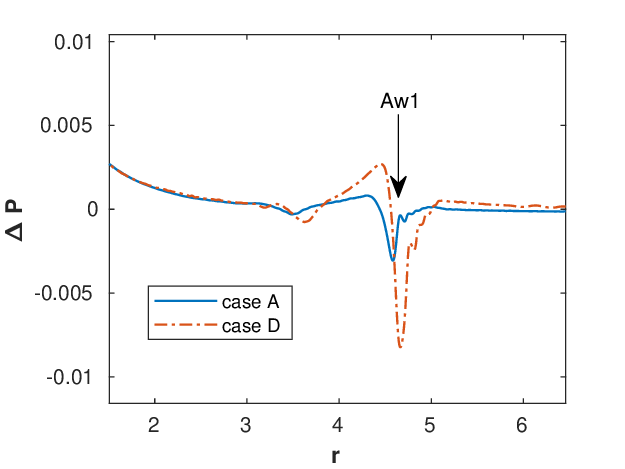}
	\caption{Pressure variation along the axis of the jet for Case A at $\bar{t} = 5.50$ and case D at $\bar{t} = 5.68$}
	\label{fig:c5_plot}
\end{figure}

\begin{figure}[h]
	\centering
	\includegraphics[width=.48\textwidth]{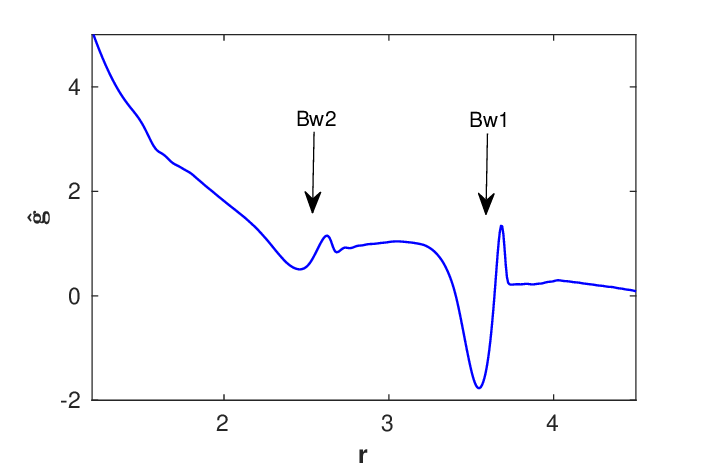}
	\caption{Case B: CB filtered variable $\hat{g}$ along the axis of the jet from time $\bar{t_1} = 4.87 \text{ to } \bar{t_n} = 6.55$}
	\label{fig:c1_plot}
\end{figure}

\begin{figure}[h]
	\centering
	\includegraphics[width=.48\textwidth]{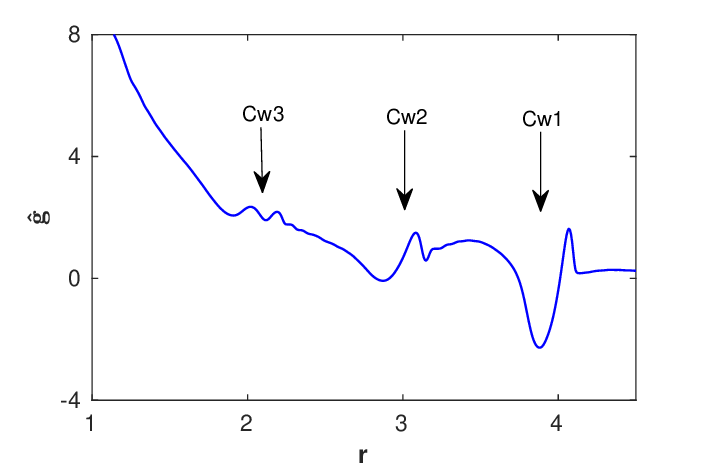}
	\caption{Case C: CB filtered variable $\hat{g}$ along the axis of the jet from time $\bar{t_1} = 5.30 \text{ to } \bar{t_n} = 6.82$}
	\label{fig:c3_plot}
\end{figure}

In case A, $M_v$ was the lowest among all the cases and produced only one pressure pulse, denoted as $Aw_1$, as shown in Figs. \ref{fig:c5_contour} and \ref{fig:c5_plot}.
As previously discussed, in case B, two pressure pulses are generated due to repetitive interaction with the vortex ring, marked as $Bw_1$ and $Bw_2$ in Figs. \ref{fig:58} and \ref{fig:c1_plot}. Although small fluctuations following $Bw_2$ and $Aw_1$ are present in Figs. \ref{fig:c5_plot} and \ref{fig:c1_plot}, they are negligible in nature.
Figs. \ref{fig:c1_plot} and \ref{fig:c3_plot} shows the CB filtered results along the axis of the jet which as discussed can be used to also represent the filtered acoustic pressure. 
The pressure pulses generated due to nested interactions in case C are shown and marked in Fig. \ref{fig:c3_plot}.
As $M_v$ is increased in case C, number pressure pulses increases by one over the previous case, as seen in Fig. \ref{fig:c3_contour} and \ref{fig:c3_plot}.
The number of pressure pulses generated in all cases are summarized in Table \ref{tab:svi} in column labelled as W.

As the vortex ring Mach number increases, the shock-vortex interaction becomes stronger as seen in Figs. \ref{fig:c5_plot},\ref{fig:c1_plot} and \ref{fig:c3_plot}. This stronger interaction results in higher number of pulse generation which is directly a consequence of the increase in nested interactions.
In stronger interactions such as in case B and C, first pressure pulse is followed by additional pressure pulses consist of compression and rarefaction due to nested interactions with vortex ring as seen in Figs. \ref{fig:c1_plot} and \ref{fig:c3_plot}. 
More number of pressure pulses are generated as number of nested increases but tend to be successively weaker in nature, which is visible in Fig. \ref{fig:c3_plot}. 
The current technique allows us to reach higher vortex Mach numbers resulting in increased in the number of nested interaction while the CBF allows a clean tagging of a pressure pulse with an unique nested interaction.

To study the effect of shock wave Mach number, case A, case D and case E are simulated at constant $M_v$ while varying the shock wave Mach number ($M_s$). Magnitude of acoustic wave increases as $M_s$ increases which can be observed in Fig. \ref{fig:c1_plot} similar to that in SVI with an analytically modelled vortex ring. When the shock wave interacts with the vortex ring, the nature of shock wave reflection changes from regular reflection to Mach reflection occurs starting from $M_s \geq 1.10$ as in the two-dimensional case\cite{chatterjee1999shock}.

\section{Conclusions}
Numerical simulations were performed, solving the compressible axisymmetric Navier-Stokes equations to study the interaction between naturally generated vortex rings by a high-speed jet pulse emanating from a circular nozzle and a moving shock wave. 
Usually, the computational study of shock wave-vortex ring interactions involve an analytically modelled vortex ring, imposed in the computational domain and can involve incompressibility assumptions in its definition. 
Generating a compressible vortex ring naturally allowed us to study vortex ring-shock interaction at higher vortex Mach numbers. 
A qualitative and quantitative comparison of the flow field due to the vortex ring-shock wave interaction was performed with literature under similar conditions of lower Mach numbers, and it showed excellent agreement. A characteristics-based filter is used to isolate acoustic and hydrodynamic fluctuations in the flow field to analyse the sound waves generated due to shock-vortex ring interaction.
This is especially useful, as the jet exit and vortex Mach number increases which also leads to higher number of acoustic waves due to more number of nested interactions. This may not be possible to be simulated in isolated SVI using analytically modelled vortex rings which are restricted to lower vortex Mach numbers.
The magnitude of the pressure wave increases with an increase in the shock wave Mach number and an increase in the vortex ring Mach number. 
In-principle, the vortex Mach number in the present study can be increased until the jet exit Mach number becomes supersonic, leading to the appearance of shocklets and shock waves in the jet. The results from this study can also be used to validate studies involving analytically imposed vortex models for SVI which may be developed for high vortex Mach numbers.

\begin{acknowledgments}
We acknowledge the use of the computing resources at ACE Facility, Department of Aerospace Engineering, IIT Bombay, India
\end{acknowledgments}

\bibliography{aipsamp}

\end{document}